\documentclass[times,5p]{elsarticle}
\usepackage{amssymb}
\usepackage{amsmath}  
\usepackage{booktabs} 
\usepackage{array} 

\usepackage{dcolumn} 
\newcolumntype{d}{D{.}{.}{-1}}
\usepackage{soul} 
\usepackage{multirow}
\usepackage{bm}
\usepackage{color}

\newcommand{\cmt}[1]{} 
\definecolor{lightgray}{rgb}{0.83, 0.83, 0.83}

\journal{Elsevier}

\usepackage{diagbox} 
\usepackage{booktabs} 
\usepackage{tikz}
\usepackage{siunitx}  

\newcolumntype{L}[1]{>{\raggedright\arraybackslash}p{#1}}

\usepackage{tabularx}
\usepackage{soul} 
\setul{1.5pt}{1.pt}  
\setstcolor{red}  

\begin{document}

\begin{frontmatter}
\title{Predicting unstable planar fault energies in multicomponent alloys from pure elemental components}

\author[1]{Chunguang Tang}
\ead{chunguang.tang@deakin.edu.au}
\author[1]{Jinqiao Liu}
\ead{jinqiao.liu@deakin.edu.au}
\author[2]{Jiaojiao Yi}
\ead{yijiaojiao2006@163.com}
\author[1]{Matthew R. Barnett}
\ead{matthew.barnett@deakin.edu.au}
\affiliation[1]{organization={Institute for Frontier Materials},
             addressline={Deakin University},
             city={Geelong},
             postcode={3217},
             state={Victoria},
             country={Australia}}
\affiliation[2]{organization={Laboratory of Advanced Multicomponent Materials, School of Mechanical Engineering},
             addressline={Jiangsu University of Technology},
             city={Changzhou},
             postcode={213001},
             state={Jiangsu},
             country={China}}

\begin{abstract}
The unstable planar fault energies (UPFEs) govern slip and twinning but are not experimentally measurable. Using EAM potentials complemented by DFT, we investigate their temperature and compositional dependence and their relationship with the shear modulus. For pure \textit{fcc} elements with low stacking fault energies (SFEs), UPFEs are substantially less temperature-sensitive than SFEs. The reported linear relationship between UPFEs and the product of shear modulus and Burgers vector ($\mu b_{\mathrm{p}}$) holds for close-packed elements and solid solutions, but deviates for intermetallics and non-close-packed elements. In the Co-Cr-Fe-Mn-Ni system, binary alloys show composition-dependent deviations, while the correlation strengthens with increasing chemical complexity. Furthermore, $\mu b_{\mathrm{p}}$ of multicomponent alloys linearly correlates with the composition-weighted elemental $\mu b_{\mathrm{p}}$ ($R^2=0.95$), despite a systematic negative deviation. This provides a simple, computationally inexpensive route for estimating $\mu b_{\mathrm{p}}$ and screening UPFEs directly from pure elemental constituents.

\end{abstract}




\end{frontmatter}


The formation of stacking faults and deformation twins plays a central role in twinning-induced plasticity and transformation-induced plasticity \cite{meyers_onset_2001, kibey_predicting_2007}. These processes are governed by the generalized stacking-fault energy (GSFE) and planar-fault energy (GPFE) landscapes, whose characteristic features include the unstable stacking-fault energy ($\gamma_{_\mathrm{USF}}$), the intrinsic stacking-fault energy ($\gamma_{_\mathrm{ISF}}$), the unstable twinning-fault energy ($\gamma_{_\mathrm{UTF}}$), and the extrinsic stacking-fault energy ($\gamma_{_\mathrm{ESF}}$). Among these, $\gamma_{_\mathrm{USF}}$ and $\gamma_{_\mathrm{UTF}}$ define the energy barriers for dislocation slip and twin nucleation, respectively. Together with $\gamma_{_\mathrm{ISF}}$, they have been widely used to characterize twinnability and deformation mechanisms \cite{bernstein_tight-binding_2004, tadmor_first-principles_2004, asaro_mechanistic_2005, cai_competition_2014}. Accurate prediction of these fault energies is therefore essential for understanding plastic deformation and for the computational design of structural alloys.

Despite their importance, the prediction of planar fault energies remains challenging. As a fundamental quantity, $\gamma_{_\mathrm{ISF}}$ has been extensively investigated both experimentally and theoretically, yet significant discrepancies between calculations and experiments persist. For example, density functional theory (DFT) calculations often predict negative $\gamma_{_\mathrm{ISF}}$ for several alloy compositions, in stark contrast to positive experimental values \cite{sun_can_2021}. These discrepancies stem from both computational and experimental limitations \cite{werner_experimental_2021}. On the computational side, finite-temperature effects, magnetic disorder, and vibrational entropy are often neglected in DFT due to their high computational cost \cite{wagner_effects_2022, lizarraga_first_2017, zhao_stacking_2017}. Moreover, $\gamma_{_\mathrm{ISF}}$ is intrinsically small and therefore highly sensitive to numerical errors. For instance, within the Ising-model framework, an error of only 1 meV/atom in the relative energies of competing crystal structures can translate into an error of 3-9 mJ/m$^2$ in the predicted stacking-fault energy \cite{tang_exploring_2025, tang_unpublished_2026}. Such uncertainties become particularly consequential for low-$\gamma_{_\mathrm{ISF}}$ materials.

\begin{figure*}[]
\includegraphics[width=7.2in]{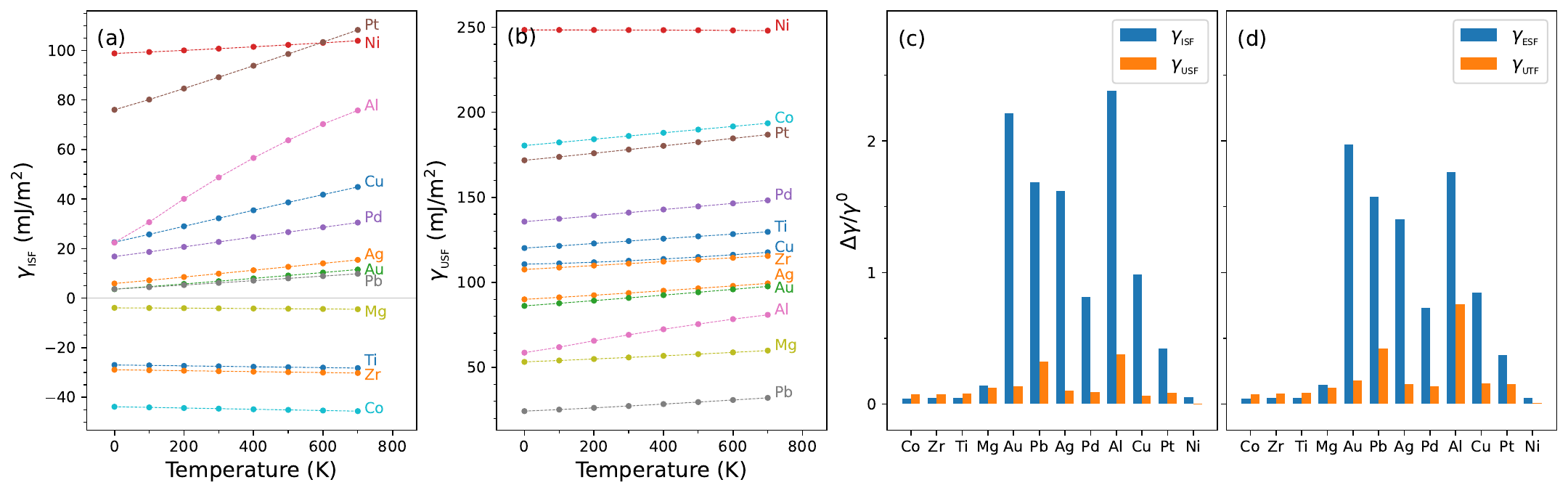}
\caption{(a-b) Temperature dependence of $\gamma_{_\mathrm{ISF}}$ and $\gamma_{_\mathrm{USF}}$. (c) The normalized changes in $\gamma_{_\mathrm{ISF}}$ and $\gamma_{_\mathrm{USF}}$, defined as $(\gamma^{700}-\gamma^{0})/\gamma^{0}$, where the superscripts are for temperature. (d) The normalized changes in $\gamma_{_\mathrm{ESF}}$ and $\gamma_{_\mathrm{UTF}}$.}
\label{fig-temp}
\end{figure*}

Although $\gamma_{_\mathrm{USF}}$ and $\gamma_{_\mathrm{UTF}}$ are generally much larger than $\gamma_{_\mathrm{ISF}}$ and thus less sensitive to small energy errors, they cannot be measured directly in experiments and must instead be obtained computationally. Efficient prediction of these quantities is particularly desirable for high-throughput alloy screening, where direct GSFE and GPFE calculations for a large number of compositions become prohibitively expensive. Using first-principles calculations, Vamsi \textit{et al.} reported a nearly linear relationship between $\gamma_{_\mathrm{USF}}$, $\gamma_{_\mathrm{UTF}}$, and $\mu b_{\mathrm{p}}$, where $\mu$ is the shear modulus and $b_{\mathrm{p}}$ is the partial Burgers vector on the \{111\} planes, for a number of pure \textit{fcc} metals \cite{vamsi_high-throughput_2021}. They further showed, using literature data, that several solid-solution alloys approximately follow the same trend. This correlation suggests that unstable fault energies may be estimated from elastic properties, providing a potentially efficient route for alloy screening. However, obtaining shear moduli from DFT still requires elastic-constant calculations, which remains computationally expensive for chemically complex alloys.

In this work, we first assess the relative sensitivity of the characteristic unstable fault energies to temperature using embedded atom method (EAM) potentials. We find that $\gamma_{_\mathrm{USF}}$ and $\gamma_{_\mathrm{UTF}}$ vary considerably less with temperature than $\gamma_{_\mathrm{ISF}}$, indicating that the unstable fault energies are substantially more robust than the intrinsic stacking-fault energy. Motivated by this robustness, we revisit the recently proposed linear relationship between unstable fault energies and shear modulus. We first validate the relationship for pure \textit{fcc} elements and then extend the analysis to various suballoys of the Cantor alloy system Co-Cr-Fe-Mn-Ni. We further show that $\mu b_{\mathrm{p}}$ for alloys can be approximated by the composition-weighted constituent elemental values. These findings clarify the applicability of shear-modulus-based descriptors for unstable fault energies and provide a simple, physically motivated route for estimating these quantities in multicomponent alloys.

To assess the effect of temperature on the fault energies, we first equilibrated the \textit{fcc} lattice of a series of elements at temperatures ranging from 100 to 700 K using NPT ensemble molecular dynamics (MD) simulations at zero pressure. For the 0 K reference, the lattice constant was determined by static relaxation at zero pressure. It should be noted that this approach captures only the temperature dependence of the fault energies through lattice expansion; thermal vibrations, electronic excitations, and magnetic contributions are not included. Nevertheless, this simplification allows us to isolate the effect of lattice expansion on the fault energies, which is the primary focus of the present temperature-dependent analysis.

For fault energy calculations, we then constructed a slab consisting of 15 (111) atomic layers, separated from its periodic images in the [111] direction by a vacuum layer of 15 \AA. Periodic boundary conditions were applied in the other two directions. For pure metals, each (111) atomic layer contains 128 atoms. For alloys, the layer size increases to up to 800 atoms, depending on the chemical complexity of the system. The GSFE was obtained by shearing the upper 8 atomic layers against the lower layers along [11$\overline{2}$] by a displacement $s_1$ up to $b_{\mathrm{p}}\!=\!a/\!\footnotesize{\sqrt{6}}$. In this scheme, $s_1\!=\!b_{\mathrm{p}}/2$ and $s_1\! = \!b_{\mathrm{p}}$ corresond to $\gamma_{_\mathrm{USF}}$ and $\gamma_{_\mathrm{ISF}}$, respectively. At $s_1\!=\!b_{\mathrm{p}}$, the bottom sheared layer was fixed, and the remaining upper layers were further sheared in the same direction by an amount $s_2$, yielding the GPFE. Here, $s_2\!=\!b_{\mathrm{p}}/2$ and $s_2\!=\!b_{\mathrm{p}}$ corresond to $\gamma_{_\mathrm{UTF}}$ and $\gamma_{_\mathrm{ESF}}$, respectively. All fault energy calculations were performed via geometry optimization at 0 K, with atomic force converged to 5 meV/\AA. For shear modulus, we computed the elastic constants by applying small strains to the equilibrium cell at 0 K. The shear modulus was then obtained from $\mu=(C_{11}-C_{12}+C_{44})/3$ \cite{vamsi_high-throughput_2021}.

All EAM-based MD simulations and static relaxation were performed using the LAMMPS code \cite{plimpton_fast_1995}. Atomic interactions were modeled using EAM potentials developed for pure metals \cite{zhou_misfit-energy-increasing_2004} and for the Cantor alloy system \cite{choi_understanding_2018}, respectively. Visualization and post-processing were carried out using the OVITO software \cite{stukowski_visualization_2009}.

\begin{figure*}[]
\includegraphics[width=7in]{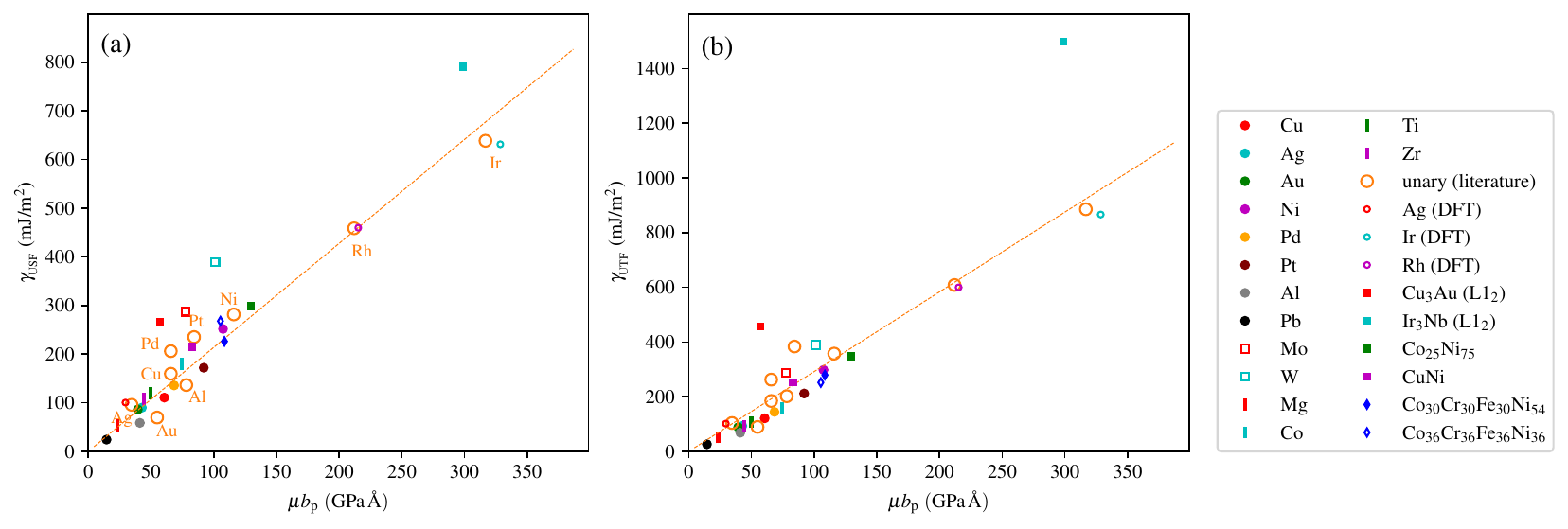}
\caption{Correlation between $\gamma_{_\mathrm{USF}}$ (a) and $\gamma_{_\mathrm{UTF}}$ (b) and $\mu b_\mathrm{p}$ for metals and alloys in \textit{fcc} structure at 0 K. The data for pure elements are EAM-based if not specified, and those for alloys are DFT-based. Note Mo and W prefer \textit{bcc} and Mg, Co, Ti, and Zr prefer \textit{hcp} at 0 K. The fitting lines and the literature data are from reference \cite{vamsi_high-throughput_2021}.}
\label{fig-pure}
\end{figure*}

For selected elements and alloys, we performed DFT calculations for comparison. In these calculations, fault energy were obtained by tilting a bulk cell consisting of 9 (111) layers, with 2 to 16 atoms per layer depending on the chemical complexity of the system. The cell tilt, applied while fixing the atomic positions, produces a relative displacement between the atoms in the cell and those in its periodic images \cite{gholizadeh_influence_2013}. The electronic relaxation was performed using the code VASP \cite{kresse_efficient_1996} with the GGA-PBE exchange-correlation functional, the projector-augmented-wave pseudopotentials as supplied in VASP, and a plane wave basis set with an energy cutoff of 500 eV. For a cubic cell about $3.5\times3.5\times3.5~$\AA$^3 $, 20$\times20\!\times\!20~k$-points were used, which corresponds to a $k$-point density of $\sim$1380 $ $\AA$^3 $ in the reciprocal space. This $k$-point density was also applied for cells of other sizes.

\begin{figure*}[h]
\includegraphics[width=6in]{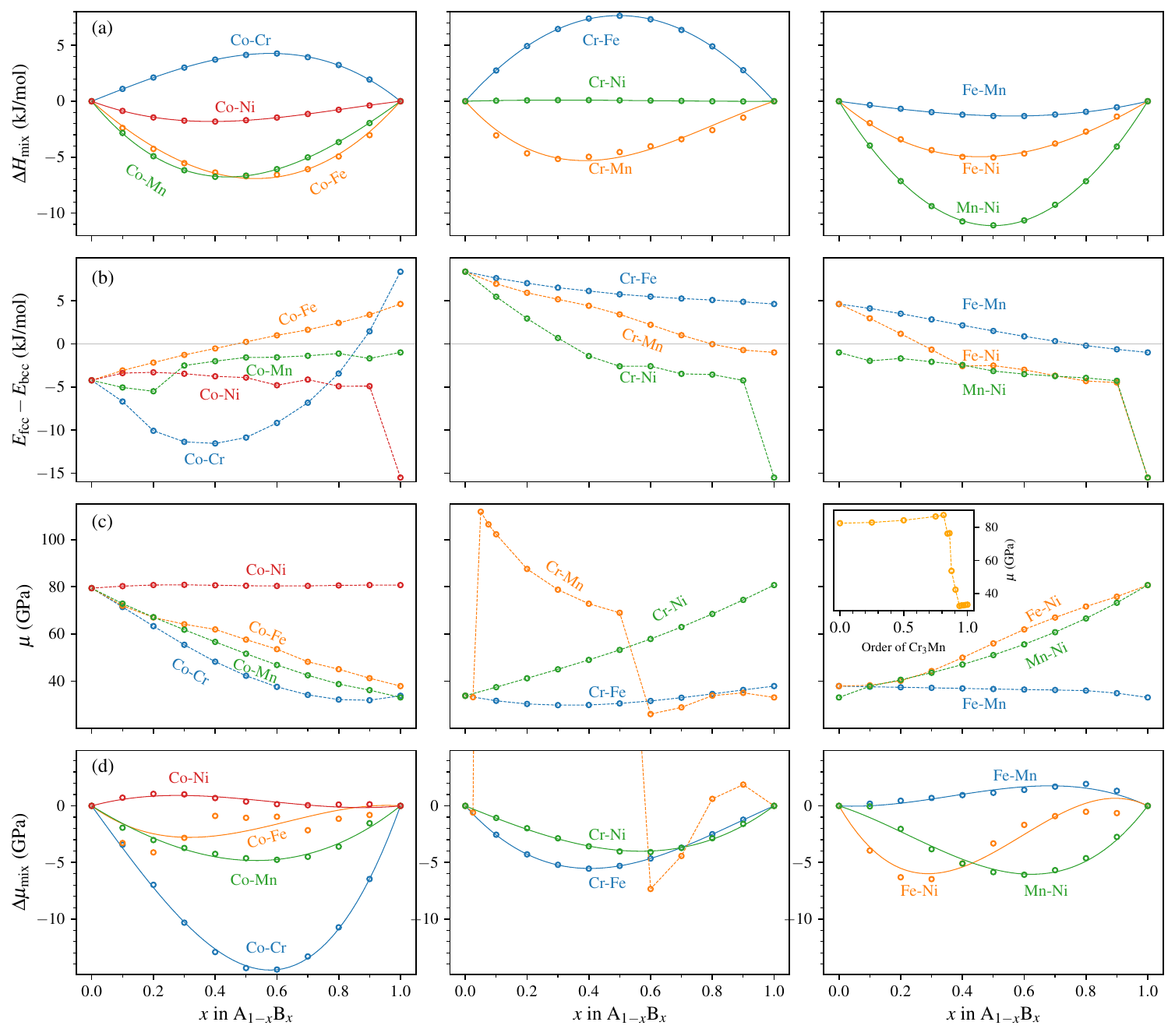}
\caption{Properties of binary sub-alloys of Co-Cr-Fe-Mn-Ni at 0 K. (a) Enthalpy of mixing $H_{\mathrm{mix}}$. The solid lines are fitting based on the Redlich-Kister model, defined as $H_{\mathrm{mix}}=(1-x)x(L_0 + L_1 (1-2x))$, where $L_0$ and $L_1$ are fitting parameters and $x$ is the atomic fraction of the second component. (b) Stability of binary alloys in \textit{fcc} phase relative to \textit{bcc}. (c) Shear modulus of the binaries. The dashed lines are for eye guide. Inset: Shear modulus of Cr$_3$Mn as a function of order parameter. (d) Effects of mixing on shear modulus plotted in a way similar to (a), with the solid lines for Redlich-Kister fitting. The shear moduli of pure elements and ordered Cr$_3$Mn (in L1$_2$ phase) are consistent with the values reported by the authors of the EAM potential \cite{choi_understanding_2018}.}
\label{fig-hmix}
\end{figure*}


Fig. \ref{fig-temp}(a) shows the temperature dependence of $\gamma_{_\mathrm{ISF}}$ for a series of pure metals in \textit{fcc} phase. Based on their 0 K values, these metals can be classified into three groups: those with low positive $\gamma_{_\mathrm{ISF}}$ (Al, Cu, Pd, Ag, Au, and Pb; below 25 mJ/m$^2$), those with high positive $\gamma_{_\mathrm{ISF}}$ (Pt and Ni; above 75 mJ/m$^2$), and those that preferentially adopt the \textit{hcp} phase (Mg, Ti, Zr, and Co), for which $\gamma_{_\mathrm{ISF}}$ is negative. With increasing temperature, the positive $\gamma_{_\mathrm{ISF}}$ values generally increase, with Pt and Al showing the most pronounced rises. In contrast, the four metastable \textit{hcp} elements exhibit slight decreases in $\gamma_{_\mathrm{ISF}}$. The corresponding temperature dependence of $\gamma_{_\mathrm{USF}}$ is shown in Fig. \ref{fig-temp}(b).

Fig. \ref{fig-temp}(c) shows the relative change in $\gamma_{_\mathrm{ISF}}$ from 0 to 700 K, normalized by the 0 K value. The metals with low positive $\gamma_{_\mathrm{ISF}}$ show the largest relative changes, whereas the other two groups exhibit only modest normalized variations. Compared with $\gamma_{_\mathrm{ISF}}$, $\gamma_{_\mathrm{USF}}$ is substantially more robust to temperature, as reflected in the normalized changes shown in Fig. \ref{fig-temp}(c). A similar temperature insensitivity was observed for $\gamma_{_\mathrm{UTF}}$ (Fig. \ref{fig-temp}(d)).

Motivated by the robustness of $\gamma_{_\mathrm{USF}}$ and $\gamma_{_\mathrm{UTF}}$, we next computed the elastic constants of the pure elements and validated their linear relationship with $\mu b_{\mathrm{p}}$, as shown in Fig. \ref{fig-pure}. Our EAM-based data are in good agreement with the DFT results of Vamsi $et~al.$ \cite{vamsi_high-throughput_2021}, with the exception of Mo and W, which favor the non-close-packed \textit{bcc} phase. Our EAM calculations at 0 K indicate that the \textit{hcp} and \textit{bcc} phases of Mo are lower in energy than the \textit{fcc} phase by approximately  1 J/mol and 15 kJ/mol, respectively. For W, the corresponding values are 1 J/mol and 16 kJ/mol.

Noting that the EAM data span smaller ranges in $\mu b_{\mathrm{p}}$ and energy compared to the DFT data, we performed additional DFT computations for selected elements (Ag, Ir, and Rh) and alloys. For pure elements and solid-solution alloys, the DFT results show good consistency with the EAM data. However, the intermetallic compounds Cu$_3$Au and Ir$_3$Nb exhibit substantial positive deviation from the established linear trend. This observation is consistent with previous theoretical and experimental studies linking increased twinning stress, $\gamma_{_\mathrm{USF}}$, and $\gamma_{_\mathrm{UTF}}$ to increased structural order \cite{chakrabortty_deformation_1977, liu_aging_2023, yang_simulation_2026}. For example, Chakrabortty, using quasichemical calculations,  reported higher twinning stress in Cu$_3$Au with increasing long-range order \cite{chakrabortty_deformation_1977}. More recently, Liu $et~al.$ showed that aging elevates chemical short-range order in CoCrNi, thereby increasing its twinning stress \cite{liu_aging_2023}. Similarly, Yang $et~al.$ found, through MD simulations, that $\gamma_{_\mathrm{USF}}$ and $\gamma_{_\mathrm{UTF}}$ in CoCrNi alloys with short-range order are elevated compared to random solid solutions \cite{yang_simulation_2026}.

Despite its breakdown for ordered structures, the linear relationship between $\mu b_{\mathrm{p}}$ and unstable fault energies may still prove useful for the design of chemically complex systems such as high-entropy alloys. To assess its applicability, we examined this relationship within the Co-Cr-Fe-Mn-Ni system using the EAM framework. We first evaluated its performance in binary alloys.

Fig. \ref{fig-hmix}(a) presents the enthalpy of mixing for the ten binary systems, which is well described by the Redlich-Kister model \cite{redlich_algebraic_1948}. To assess the impact of phase stability, we computed the relative energies of the \textit{fcc} binaries against their \textit{bcc} counterparts. As shown in Fig. \ref{fig-hmix}(b), the Cr-Fe alloy favors the \textit{bcc} phase across the composition range, whereas most other binaries exhibit a partial preference for \textit{bcc}.

Fig. \ref{fig-hmix}(c) shows the composition-dependent shear modulus of the same binaries. Most binaries exhibit a smooth, continuous variation with composition. A notable exception is the Cr-Mn system, where the shear modulus shows an abrupt and substantial increase for Mn concentrations between 5 and 50 at.\%. To understand this anomaly, we computed the shear modulus of Cr$_3$Mn as a function of order parameter $s$, defined as $s=(r - c)/(1-c)$, where $c$ is the Mn concentration and $r$ is the fraction of Mn atoms occupying the Mn sublattice (inset of Fig. \ref{fig-hmix}(c)). In the fully ordered L1$_2$ structure ($s=1$), Cr$_3$Mn exhibits a shear modulus of 33 GPa, consistent with the literature values \cite{choi_understanding_2018}. However, as $s$ deceases to 0.8, achieved by randomly swapping Cr and Mn atoms in the L1$_2$ structure, the shear modulus rises sharply to approximately 80 GPa. This pronounced sensitivity suggests that the EAM potential captures a strong coupling between chemical ordering and elastic properties in the Cr–Mn system, a feature that warrants further scrutiny. For the remaining binaries, the mixing effect on the shear modulus can be well approximated by the Redlich–Kister model (Fig. \ref{fig-hmix}(d)), mirroring the behavior observed for the enthalpy of mixing.

\begin{figure*}[]
\includegraphics[width=6in]{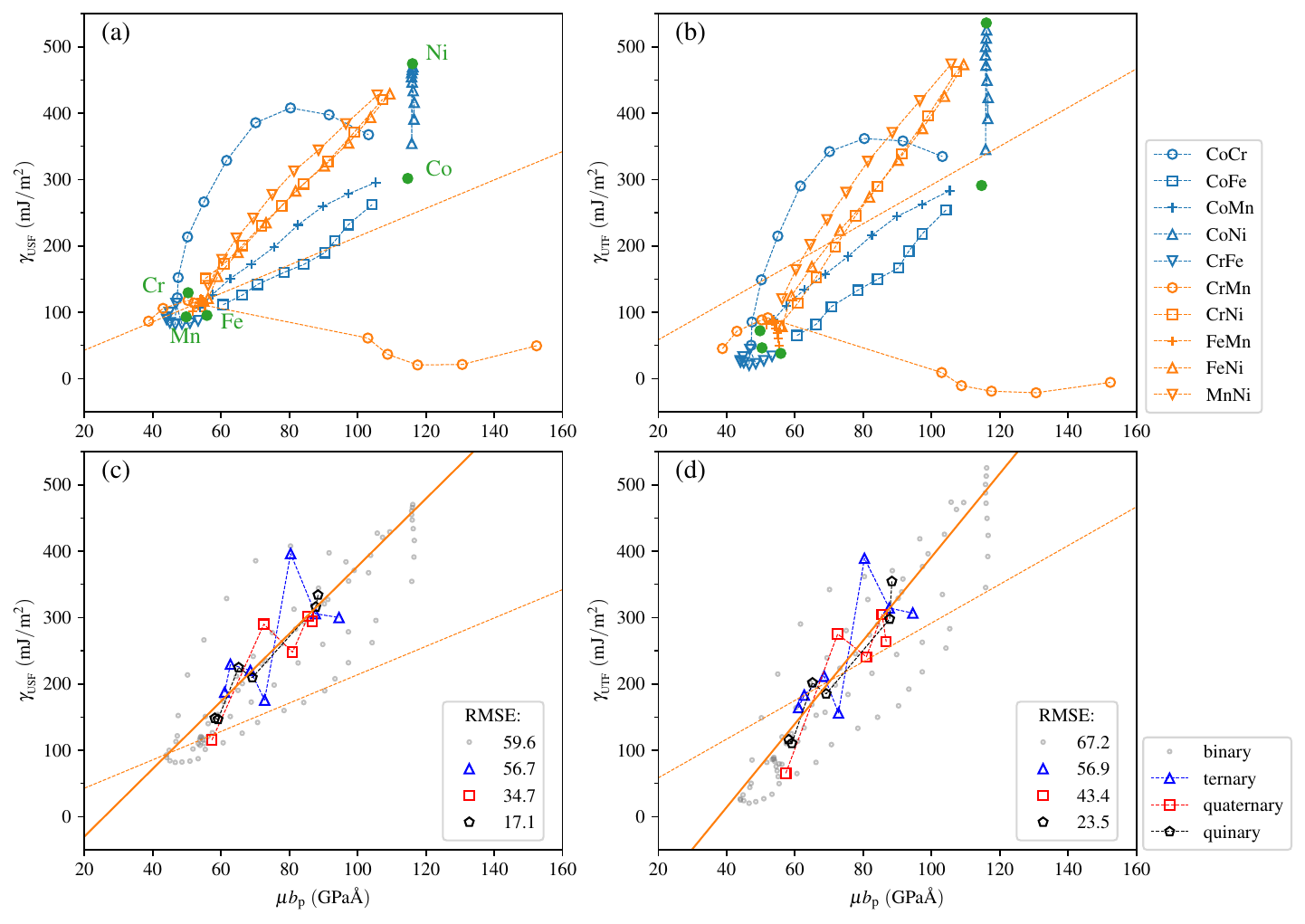}
\caption{(a-b) Correlation between $\gamma_{_\mathrm{USF}}$, $\gamma_{_\mathrm{UTF}}$ and $\mu b_\mathrm{p}$ for pure components and binary alloys in the Co-Cr-Fe-Mn-Ni system. The curved dashed lines are for eye guide. (c-d) The correlation for binary and higher-order alloys in the Co-Cr-Fe-Mn-Ni system, restricted to compositions with Cr$+$Mn$<$50 at.\%. The reported root-mean-square error (RMSE) values correspond to the deviations from the linear fits. The solid lines represent fits to the ternary and higher-order alloy data, while the straight dashed lines in all figures are reproduced from reference \cite{vamsi_high-throughput_2021}.}
\label{fig-hmu}
\end{figure*}

\begin{figure*}[]
\includegraphics[width=7in]{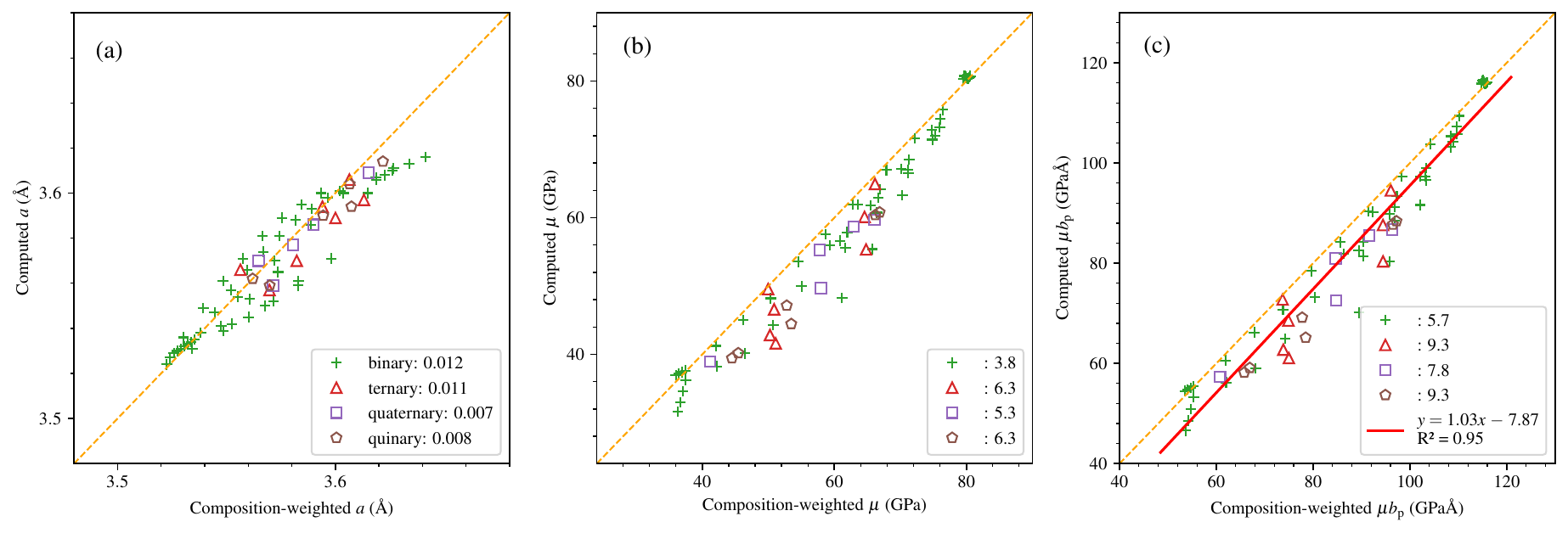}
\caption{Comparison of the computed and composition-weighted values for (a) lattice constants, (b) shear moduli, and (c) $\mu b_{\mathrm{p}}$, with the corresponding RMSE values indicated. The composition-weighted lattice constant is calculated from the weighted cell volume. The solid line in (c) is a linear fit to all data points. The alloy compositions are restricted to Cr$+$Mn$<$50 at.\%}
\label{fig-linear}
\end{figure*}

We next computed the unstable fault energies of the binaries together with those of their pure components. As shown in Fig. \ref{fig-hmu}(a-b), the binaries exhibit diverse trends and do not follow the universal linear relationship between the unstable fault energies and $\mu b_{\mathrm{p}}$ observed in Fig. \ref{fig-pure}. Several factors may contribute to these deviations. First, the pronounced increase in $\mu$ for Cr-Mn causes this system to deviate substantially from the trends of the other binary systems. Second, the present EAM potential predicts substantially higher unstable fault energies for Ni than the corresponding DFT values. Consequently, all Ni-containing binaries progressively deviate from the DFT-derived linear relationship as their compositions approach the Ni-rich end. We also note that Co-Cr exhibits a more pronounced curvature than the other binary systems, consistent with its relatively larger absolute values of $\Delta\mu_{\mathrm{mix}}$.

We observed no clear correlation between these diverse $\mu b_{\mathrm{p}}-\gamma$ trends and the relative stability of the \textit{fcc} against the \textit{bcc} phase. For the Cr-Fe system, which remains \textit{bcc}-favored across the entire composition range, both the $\mu b_{\mathrm{p}}$ and $\gamma$ values span narrow ranges, rendering its overall trend less pronounced than those of the other binaries. Furthermore, for systems exhibiting partial \textit{bcc} stability, the \textit{bcc}-favored composition ranges showed no significant deviations in behavior when compared to their \textit{fcc}-favored counterparts. A plausible explanation is that the substantial energy barrier for the \textit{fcc}-to-\textit{bcc} transformation effectively preserves the metastability of the \textit{fcc} phase, preventing phase stability from directly influencing the unstable fault energies.

The weaker linearity observed in the binary alloys indicates that the relationship between unstable fault energies and $\mu b_p$ is sensitive to alloy chemistry. The specific behavior may also depend on the interatomic potential, particularly because different potentials can be optimized for different compositions and material properties. Nevertheless, as shown in Fig. \ref{fig-hmu}(c-d), the relationship becomes progressively more linear as the number of components increases. This systematic trend suggests that increasing chemical complexity may enhance the transferability of the $\gamma_{_\mathrm{USF}}$-$\mu b_\mathrm{p}$ and $\gamma_{_\mathrm{UTF}}$-$\mu b_\mathrm{p}$ relationships. (We note, however, that the potential was developed for multicomponent alloys, which may contribute to this trend).

It is worth noting that the slope obtained for the Co-Cr-Fe-Mn-Ni system differs from that established for the pure close-packed elements and solid-solution alloys considered above. This difference may arise, at least in part, from the use of different EAM potentials in the respective datasets. In particular, the EAM potential employed for the Cantor alloy substantially overestimates the unstable fault energies of Ni compared with both the other EAM and DFT calculations, which can shift the $\gamma_{_\mathrm{USF}}$-$\mu b_{\mathrm{p}}$ and $\gamma_{_\mathrm{UTF}}$-$\mu b_{\mathrm{p}}$ relationships and hence alter their slopes. Therefore, the slope of the correlation should not necessarily be regarded as universal across different interatomic potentials. Importantly, however, within a given EAM potential the relationship remains internally self-consistent across the corresponding elemental and alloy systems. The progressively improved correlation observed from binary to ternary, quaternary, and quinary Cantor alloys therefore remains useful for screening compositions within the same potential framework, even if its slope differs from that obtained using other potentials or DFT.

The correlation between the unstable fault energies and $\mu b_{\mathrm{p}}$ provides a potentially useful approach for screening alloy compositions with desirable $\gamma_{_\mathrm{USF}}$ and $\gamma_{_\mathrm{UTF}}$. However, obtaining $\mu b_{\mathrm{p}}$ for a chemically complex alloy can itself be computationally demanding, particularly when first-principles calculations are required to determine the elastic constants. The Redlich-Kister fitting of the mixing effects on the shear modulus (Fig. \ref{fig-hmix}(c)), together with privious CALPHAD-based models for composition-dependent elastic properties \cite{ marker_elastic_2017, bocklund_implementation_2024}, suggests a possible route for estimating alloy elastic properties within a CALPHAD framework. However, such an approach requires interaction parameters whose accuracy and transferability are not guaranteed, particularly for multicomponent systems \cite{tang_searching_2026}. This motivates us to explore whether $\mu b_{\mathrm{p}}$ can instead be estimated directly from the properties of the pure constituent elements, providing a simple and computationally inexpensive route for screening multicomponent alloys.

We first examined whether the lattice constants and shear moduli of the alloys could be approximated from the corresponding properties of their pure constituent elements. The composition-weighted lattice constant was defined as $a_{\mathrm{mix}}=\left(\sum_i c_i v_i\right)^{1/3}$, where $c_i$ and $v_i$ are the concentration and cell volume of component $i$, respectively. As shown in Fig. \ref{fig-linear}(a), the calculated lattice constants closely follow the $y=x$ line, with small RMSE values, indicating that the composition-weighted estimate provides an excellent approximation across the different alloy compositions. For the shear modulus, the calculated values also show a strong correlation with the composition-weighted elemental values, although they systematically fall slightly below the $y=x$ line (Fig. \ref{fig-linear}(b)). The RMSE ranges from 3.8 to 6.3 GPa depending on the number of components. Since $b_{\mathrm{p}}$ is determined by the lattice constant, combining the composition-weighted estimates of the lattice constant and shear modulus naturally leads to a similar correlation for $\mu b_{\mathrm{p}}$. As shown in Fig. \ref{fig-linear}(c), the calculated $\mu b_{\mathrm{p}}$ values are well described by a linear relationship with the composition-weighted values, $y=1.03x-7.87$, with $R^2=0.95$. These results demonstrate that the $\mu b_{\mathrm{p}}$ required for estimating unstable fault energies can be obtained with reasonable accuracy directly from the properties of the pure constituent elements.


In this work, we investigated the temperature dependence and compositional trends of generalized planar-fault energies, as well as their relationship with the shear modulus, in \textit{fcc} metals and alloys using EAM potentials. For a series of pure elements with relatively small positive $\gamma_{_\mathrm{ISF}}$ ($\leq\!25$ mJ/m$^2$), $\gamma_{_\mathrm{ISF}}$ exhibits substantially greater relative variation with temperature than $\gamma_{_\mathrm{USF}}$ and $\gamma_{_\mathrm{UTF}}$, indicating that the unstable fault energies are less sensitive to temperature. We further examined the previously reported correlation between the unstable fault energies and $\mu b_{\mathrm{p}}$. For a series of pure close-packed elements and solid-solution alloys, our EAM and DFT calculations yield a clear linear relationship, with a slope consistent with the literature. Substantial deviations are observed for intermetallic compounds and the non-close-packed elements Mo and W. We then investigate the Cantor alloy system Co-Cr-Fe-Mn-Ni, where EAM calculations reveal composition-dependent trends in binary alloys. In contrast, the relationship becomes progressively more linear as the chemical complexity increases from binary to ternary, quaternary, and quinary alloys.

To facilitate the use of $\mu b_{\mathrm{p}}$ as a descriptor for alloy screening, we explored whether this quantity can be estimated directly from the properties of the pure components. The alloy lattice constants are accurately described by taking the cube root of the composition-weighted cell volume, while the calculated shear moduli show a strong positive correlation with the corresponding composition-weighted elemental values. The latter systematically overestimate the alloy shear moduli, indicating a non-negligible contribution from chemical interactions. Combining the estimates of the two yields a similarly strong correlation for $\mu b_{\mathrm{p}}$, which can be described by a simple linear relationship with $R^2=0.95$.

These results demonstrate that $\mu b_{\mathrm{p}}$ can be estimated with reasonable accuracy from pure-component properties without explicitly calculating the elastic constants of each alloy composition. Together with the established correlation between $\mu b_{\mathrm{p}}$ and the unstable fault energies, this provides a simple and computationally inexpensive approach for screening $\gamma_{_\mathrm{USF}}$ and $\gamma_{_\mathrm{UTF}}$ across multicomponent alloy compositions.

\noindent{\bf{Declaration of competing interest}}
The authors declare that they have no known competing financial interests or personal relationships that could have appeared to influence the work reported in this paper.

\noindent{\bf{Data availability}}
Data will be made available on request.

\noindent{\bf{Acknowledgements}}
The ARC funding (FL210100147) is gratefully acknowledged.

\end{document}